\documentclass[fleqn,a4paper,12pt]{article}
\usepackage{amssymb}
\usepackage{makeidx}
\usepackage{amsmath}
\usepackage{graphicx}
\usepackage{lscape}
\usepackage{a4}
\usepackage{epsfig}
\begin{document}

\thispagestyle{empty}
\newcommand{\al}{\alpha}
\newcommand{\bt}{\beta}
\newcommand{\s}{\sigma}
\newcommand{\lbd}{\lambda}
\newcommand{\vp}{\varphi}
\newcommand{\va}{\varepsilon}
\newcommand{\gm}{\gamma}
\newcommand{\G}{\Gamma}
\newcommand{\p}{\partial}
\newcommand{\om}{\omega}
\newcommand{\be} {\begin{equation}}
\newcommand{\lo}{\left(}
\newcommand{\ro} {\right)}
\newcommand{\ee} {\end{equation}}
\newcommand{\ba} {\begin{array}}
\newcommand{\ea} {\end{array}}
\newcommand{\ds}{\displaystyle}
\begin{center}
{\large\bf Group classification of systems of non-linear
reaction-diffusion equations with general diffusion matrix.
III. Triangular diffusion matrix }
\end{center}
\vspace{2mm}
\begin{center}
  A. G. Nikitin\\

Institute of Mathematics of Nat.Acad. Sci of Ukraine, 4
Tereshchenkivska str. 01601 Kyiv, Ukraine
\end{center}
 \vspace{2mm}

{\bf Absrtract.} Group classification of systems of two coupled nonlinear
reaction-diffusion equation with a general diffusion matrix started in papers
\cite{N1}, \cite{N2} is completed in present paper where all non-equivalent equations
with triangular diffusion matrix are classified.
 In addition, symmetries of diffusion systems with nilpotent diffusion matrix
and additional first order derivative terms are described.

\vspace{2mm}

\section{Introduction}

Group classification can be considered as one of basic elements of qualitative analysis of
entire classes of differential equations. Such classification makes it possible to select
the basic non-equivalent equations in the considered class of them. In addition, it opens the
way to application of powerful  group-theoretical tools for searching for conservation laws,
construction of exact solutions, group generation of families of solutions starting with
known ones, etc., etc.

In the present paper we complete the group classification of systems
of reaction-diffusion equations with general diffusion matrix performed in papers
\cite{N1}, \cite{N2}. These equations can be written in the following form:
    \be
\label{1.1}\ba{l}
\displaystyle u_t-\Delta(A^{11} u+A^{12} v)=f^1(u,v),\\
    \displaystyle v_ t-\Delta(A^{21} u+A^{22} v)=f^2(u,v) \ea \ee
where $u$ and $v$ are function of $t, x_1, x_2,  \ldots , x_m$,
 $\ A^{11}, A^{12}, A^{21}$ and $A^{22}$ are real constants
and $\Delta$ is the Laplace operator in $R^m$.

Equations (\ref{1.1}) form the grounds for a great many models of mathematical physics,
biology, chemistry, etc., which makes their group classification to be especially
relevant. We will not discuss here the history of investigation of symmetries
of systems (\ref{1.1}) which can be found in \cite{N1}, \cite{N2}.

Up to linear transformation of dependent variables there exist
three {\it a priori} non-equivalent versions of equations
(\ref{1.1}), corresponding to a diagonal, triangular and square
diffusion matrix. The equations with diagonal and square diffusion
matrix have been studied in papers \cite{N1} and \cite{N2}
correspondingly were the complete group classification of such
equations is carried out.

Here we consider the remaining case when the diffusion matrix is triangular and system
(\ref{1.1}) is reduced to the following form:
\be\label{1.2}\ba{l}
\displaystyle u_t-a\Delta u=f^1(u,v),\\
    \displaystyle v_ t-\Delta u-a\Delta  v=f^2(u,v) \ea \ee
    where $a$ is a real constant.

In the case when $a=0$ the diffusion matrix is nilpotent and it is naturally to
generalize (\ref{1.2}) to the following system
\be\label{1.3} \ba{l}
u_t-p_\mu v_{x_\mu}=f^1(u_1, u_2), \\
v_t-\Delta u=f^2(u_1, u_2)
 \ea \ee
 were (and in the following text) summation from 1 to $m$ is imposed over the repeated Latin
 indices, $p_\mu$ are arbitrary constants. Using linear transformations of independent
 variables we reduce values of $p_\mu$ to \be\label{1.31} p_1=p_2=\cdots=p_{m-1}=0, \
p_m=p\ee
where $p=\sqrt{p_1^2+p_2^2+\cdots+p_m^2}$.

We notice that in addition to possible application in mathematical biology equations
(\ref{1.3}) can serve as a
potential equation for the nonlinear D'alembert equation.

\section{Symmetries and determining equations}

To describe  symmetries of equations (\ref{1.2}) and (\ref{1.3})
with respect to continuous groups of transformations we use the Lie
infinitesimal approach.
Using the standard Lie algorithm we can find
the determining
equations for coefficient functions $\eta ,\ \xi _{a},\ \pi^b$ of generator $X$
of the symmetry group:
\begin{equation}
X=\eta {\partial_ t}+\xi ^{\nu}{\partial_{
x_{\nu}}}-\pi ^{1}{\partial_{ u}} -\pi ^{2}{\partial_{ v}} \label{3.105}\ee
The first set of determining equations describes
dependence of $\eta, \xi^\nu$ and $\pi^b$ on $u$ and $v$:
\begin{equation}
 \eta_{ u}=\eta_v=0,\ \xi ^{\nu}_{u}=\xi ^{\nu}_{v}=0,\ \ \pi ^{a}_{uu}=\pi ^{a}_{uv}
 =\pi ^{a}_{vv}=0. \label{3.4a}
\end{equation}
So from (\ref{3.4a}) $\eta $ and $\xi ^{\nu}$ are functions of $t$ and $x_{\mu}$,
and
\begin{equation}
\pi ^{a}=N^{a1}u+N^{a2}v+B ^{a},\ \  a=1,2  \label{3.5a}
\end{equation}
where $N^{ab}, B^a$ are functions of $t $ and $x_{\nu}$ only.
The remaining determining equations in the case $a\neq 0$ are \cite{N1}:
\begin{equation}
2A\xi _{x_\mu}^{\nu}=-\delta ^{\mu\nu}(\eta_t A+[A,N ]),\qquad {\eta}
_{x_\nu t}=0, \label{4.2a}
\end{equation}
\begin{equation}\ba{l}
{\xi}^\nu_t-2aN^{11}_{x_\nu}-a\Delta\xi^{\nu}=0,\ \ N^{12}=0, \
N^{21}_{x_\nu}=-aN^{11}_{x_\nu},\\
\eta_tf^k+N^{kb}f^{b}+(N^{kb}_t-\Delta A^{ks}N^{sb})u_b
+B^k_t-\Delta A^{kc}B^c\\
=(B ^{a}+N^{ab}u_{b})f^{k}_{u_a}.\ea
\label{3.8a}
\end{equation}
Here $N$ and $A$ are matrices whose elements are $N^{ab}$ and $A^{ab}$, $\delta^{ab}$ is the
Kronecker symbol, and we use the temporary notation $u=u_1, v=u_2$.

Using (\ref{3.5a})--(\ref{3.8a}) we find
the general form of symmetry
(\ref{3.105}) admitted by equation (\ref{1.2}):
\be \label{2.4} \ba{l}
X=\Psi^{\mu \nu} x_\mu \p_{x_\nu}+\nu \p_t+\rho_\mu\p_{x_\mu}+\lbd K+\sigma_\mu G_\mu+
\om_\mu \hat G_\mu+\mu
D\\-C^{1}(u\p_u+v\p_v)-C^2u\p_v-B^1{\p_{ u}}-B^2\p_v
\ea
\ee where the Greek letters denote arbitrary constants, $B^1,\ B^2$ are
functions of $t,x$, and $C^{1},\ C^2$ are functions of $t$,
 \be
\label{2.6} \ba{l} K=2t(t\p_t+x_\mu
\p_{x_\mu})-\frac{x^2}{2}\lo\frac{1}{a}(u\p_u+v\p_v)-u\p_v\ro
-tm(u\p_u+v\p_v)
,\\
G_\mu=t\p_{x_\mu}+\frac{1}{2}x_\mu\lo\frac{1}{a}(u\p_u+v\p_v)-\frac1{a^2}u\p_v\ro,\\
\hat G_\mu=e^{\gm t}\left(\p_{x_\mu}+\frac{1}{2}\gm
x_\mu\lo\frac{1}{a}(u\p_u+v\p_v)-\frac1{a^2}u\p_v\ro\ro,\\
D=t\p_t+\frac12 x_\mu \p_{x_\mu}. \ea \ee

For $a=0$ symmetry $X$ again has the form (\ref{2.4}) where however
    $\lambda=\s_\mu=\om_\mu=0$. In addition, $B^2$ can depend not on $t,x$ only, but also on $u$.

    In accordance with (\ref{3.105}), (\ref{3.5a}) and (\ref{3.8a})
    equation (\ref{1.2}) admits symmetry (\ref{2.4}) iff the following classifying equations
are satisfied: \be \label{2.7} \ba{l} \left(\lbd(m+4
)t+\mu+\frac1a\lo\frac{1}{2}\lbd x^2+\sigma_\mu x_\mu+\gm e^{\gm
t}
\om_\mu x_\mu\ro)+C^{1}\right) f^1\\
+C^{1}_t u +B_t^1-a\Delta B^1
=\lo B^1\p_u+B^2\p_v+C^{1}(u\p_ u+v\p_v) +C^2u\p_v\right.\\\left.+\lbd mt(u\p_u+v\p_v)
+\left(\frac12\lbd x^2+\sigma_\mu x_\mu
+\gm e^{\gm t}\om_\mu
x_\mu\right) \left(\frac1a(u\p_u+v\p_v)-\frac1{a^2}u\p_v\ro\ro f^1,\\\\
\lo\lbd(m+4 )t+\mu+C^1\ro f^2+C^2f^1+\lo\frac{1}{2}\lbd
x^2+\sigma_\mu x_\mu+\gm e^{\gm t} \om_\mu
x_\mu\ro\lo\frac{1}{a}f^2-f^1\ro\\+C^1_tv+C^{2}_t u
 +B_t^2-\Delta B^2-a\Delta B^1
=\lo B^1\p_u+B^2\p_v+C^{1}(u\p_ u+v\p_v)
+C^2u\p_v\right.\\\left.+\lbd mt(u\p_u+v\p_v) +\left(\frac12\lbd
x^2+\sigma_\mu x_\mu +\gm e^{\gm t}\om_\mu x_\mu\right)
\left(\frac1a(u\p_u+v\p_v)-\frac1{a^2}u\p_v\ro\ro f^2.
 \ea \ee

 Equations (\ref{2.7}) are nothing but a special case of the generic classifying equations
(2.9), ref. \cite{N1} which are valid for arbitrary invertible diffusion matrix.

Equation (\ref{1.3})  needs a particular analysis. For $p\neq 0$
the related symmetry operator (\ref{3.105}) reduces to the form
\be\label{n001}X=X_0+\tilde X\ee where
\[X_0=\lbd\p_t+\nu_\al\p_{x_\al}+\Psi^{kl}x_k\p_l,\]
\be \label{2.11}\tilde X=\mu\left(3t\p_t+2x_\nu \p_\nu-v\p_v
\right)-F\left(u {\p_u}+v\p_v\right)-B^1{\p_ u}-B^2\p_v .\ee

Here $\Psi^{\mu\nu}$ is an antisymmetric tensor and summation is
imposed over the repeated Greek and Latin indices indices from 1
to $m$ and from 1 to $m-1$ correspondingly.

Symmetries $X_0$ are admitted by equations (\ref{1.3}),
(\ref{1.31}) with arbitrary non-linearities $f^1$ and $f^2$ while
the classifying equations generated by symmetries $\tilde X$ are
\be \label{2.12} \ba{l} (3\mu+F)f^1+F_t
u+B^1_t-pB^2_{x_m}\\=\left(B^1{\p_u}+B^2{\p_v}+ F u\p_u+(F+\mu)
v\p_v\right)f^1,\\\\
(4\mu+F)f^2+F_t v+B^2_t-\Delta B^1\\=\left(B^1\p_u+ B^2 \p_v+F
u\p_u+(F+\mu)v\p_v\right)f^2 \ea \ee where $F$ and $B^1$, $B^2$
are unknown functions of $t$ and $t, x$ respectively.

The determining equations for symmetries of equation (\ref{1.3})
with $p=0$ are qualitatively different for the cases, when the
number $m$ of spatial variables $x_1,x_2, \cdots x_m$ is $m=1,\
m=2$ and $m>2$. The related  generator (\ref{3.105}) has the form
\be\label{new}\ba{l}X=\al D+\lo\int (N-M)dt\ro{\p_ t}+2mH^a\p_{
x_a} -\lo  N+({m-2})H^a_{x_a}\ro u\p_u\\-\lo
M+{(m+2)}H^\al_{x_\al}\ro v\p_v- B^1{\p_
u}-B^2{\p_v}-B^3u\p_v\ea\ee where summation from $1$ to $m$ is
imposed over repeating indices, the Greek letters denote arbitrary
parameters, $M, N$ are functions of $t$, $B^1, B^3$ are functions
of $t, x$, $B^3$ are functions of $t,x, u$,  and
\[H^a=2\lbd_bx_bx_a-x^2\lbd_a \  \text{if}\ m>2.\]

For $m=2\ $ $H^a $ are arbitrary
functions satisfying the Caushy-Rieman conditions:
\[ H^1_{ x_1}= H^2_{x_2},\ \
 H^1_{x_2}
=- H^2_{x_1}.\]

In the case when $m=1\ \ H^1$ is a function of $x$ and the sums
with respect to $a$ in (\ref{new}) are reduced to the only terms.

The related classifying equations have the form
\be\label{last}\ba{l}\lo\al+2N-M+(m-2)H^a_{x_a}\ro f^1+N_tu+B^1_t
=\lo B^1{}\p_{u}+B^2{}\p_{v}\right.\\\left.+B^3u\p_{v}+\lo
N+(m-2)H^a_{x_a}\ro u{}\p_{u}+\lo
M+(m+2)H^a_{x_a}\ro v\p_{v}\ro f^1,\\
\\
\lo \al+N+(m+2)H^a_{x_a}\ro f^2+B^3f^1+M_tv+B^3_t u+B^2_t-\Delta
B_1\\+(2-m)\Delta H^a_{x_a}u =\lo
B^1{}\p_{u}+B^2{}\p_{v}+B^3u\p_{v}+\lo N+(m-2)H^a_{x_a}\ro
u\p_{u}\right.\\\left.+\lo M+(m+2)H^a_{x_a}\ro v\p_{v}\ro
f^2.\ea\ee

We notice that in this case symmetry classification appears to be
rather complicated and cumbersome. Nevertheless, the classifying
equations can be effectively solved using the approach outlined in
the following sections.
\section{Classification of symmetries}

Following \cite{N1} we specify basic, main and extended symmetries for the analyzed systems of
reaction-diffusion equations.

Basic symmetries form the kernel of the main symmetry group and so
are admitted by equation (\ref{1.2}) with arbitrary
non-linearities $f^1$ and $f^2$. They are generated by the
following infinitesimal operators: \be \label{4.1} P_0=\p_t, \quad
P_\lbd=\p_{x_\lbd}, \quad J_{\mu \nu}=x_\mu\p_{x_\nu}-x_\nu
\p_{x_\mu} \ee and correspond to shifts of independent variables
and rotations of variables $x_\nu$.

Main symmetries form an important subclass of general symmetries
(\ref{2.4}) which correspond to $\lbd=\s_\nu=\omega_\nu=0$ and so
have the following form \be \label{4.2} \tilde X=\mu
D-C^{1}(u\p_u+v\p_v)-C^2u\p_v-B^1\p_{ u}-B^2\p_v. \ee

To describe all Lie symmetries admitted by equation (\ref{1.2}) we follow the procedure
outlined in \cite{N1} which includes the following steps:
\begin{itemize}
\item  Finding all main symmetries (\ref{4.2}), i.e., solving equations (\ref{2.7}) for
$\Psi^{\mu\nu}=\nu=\rho_\nu=\sigma_\nu=\om_\nu=0$:
 \be \label{4.3}\ba{l} (\mu
+C^{1})f^1+C^{1}_t u_1+B^1_t-a\Delta B^1\\=
\left(C^{1}(u\p_u+v\p_v)+C^2u\p_v+B^1\p_u+B^2\p_v\right) f^1,\\\\
(\mu+C^{1})f^2+C^2_tu+C^{1}_t v+B^2_t-a\Delta B^2-\Delta B^1\\=
\left(C^{1}(u\p_u+v\p_v)+C^2u\p_v+B^1\p_u+B^2\p_v\right) f^2.\ea
\ee \item  Specifying all cases when the main symmetries can be
extended, i.e., at least one of the following systems is
satisfied: \be \label{4.8}\ba{l}
af^1=\left(a(u\p_u+v\p_v)-u\p_v\right)f^1,\\
af^2-f^1=\left(a(u\p_u+v\p_v) -u\p_v\ro f^2 ;\ea\ee
 \be \label{4.10}\ba{l} a(f^1+\gm
u)=\lo a(u\p_u+v\p_v)-u\p_v\ro f^1,\\ a( f^2+\gm
v)-\gm u=\lo a(u\p_u+v\p_v)-u\p_v\ro f^2\ea \ee
or if equation (\ref{4.8}) is satisfied together with the following condition:
 \be\label{4.9}\ba{l}(m+4)f^a=m(u\p_u+v\p_v) f^a,\ \ a=1,2
\ea\ee

If relations (\ref{4.8}), (\ref{4.10}) or (\ref{4.9}) are valid then the system (\ref{1.2})
admits symmetry $G_\al, \widehat G_\al$ or $K$ correspondingly.

\item When classifying equations (\ref{1.3}) or (\ref{1.2}) with $a=0$
the second step in not needed in as much as in accordance with (\ref{new}) and
(\ref{2.11} ) these equations admit basic and main symmetries only.
\end{itemize}

We note that the first step of the described procedure appears to be very complicated.
In spite of that equations (\ref{4.3}) can be effectively solved using separation of
variables, the tree of versions of such separations is rather large and includes a
lot of intersections.

In the next section we present specific tools used to overcome these difficulties.

\section{Algebras of main symmetries for equation (\ref{1.2})}

In accordance with the plane outlined in Section 4, to make symmetry classification of
equations (\ref{1.2}) we
first describe the main symmetries generated by operators
(\ref{4.2}) and then indicate extensions of these symmetries.

First we note that for any $f^1$ and $f^2$ equation (\ref{1.2}) admits the following
equivalence transformations
  \be \label{x.2} \ba{l}
u\to K^{1}u+b^1, \quad v\to K^1 v+K^2 u+b^2, \\f^1 \to \lbd^2
K^{1}f^1,
\quad f^2\to \lbd^2(K^1f^2+K^2f^1),\\
t \to \lbd^{-2} t, \quad x_b \to \lbd^{-1}x_b
 \ea \ee
where $K^1, K^2$ and $\lambda$ are constants which are distinct
from zero, $b^1$ and $b^2$ are arbitrary constants. In accordance
with its definition, equivalence transformations keep the general
form of equation (\ref{1.2}) but can change the concrete
realization of their r.h.s..  For some non-linearities  $f^1$ and
$f^2$ there exist additional equivalence transformation which will
be specified in the following.

We will use transformations (\ref{x.2}) to simplify generators (\ref{4.2}).

To solve rather complicated classifying equations (\ref{4.3}),
 we use the main algebraic property of
the main symmetries, i.e., the fact that they should form a Lie
algebra (which we denote by $\cal A$). In other words, instead of going
throw all non-equivalent
possibilities arising via separation of variables in the
classifying equations we first specify all non-equivalent
realizations of algebra $\cal A$ for our equations up to
arbitrary constants and arbitrary functions. Then we easily solve
classifying equations (\ref{4.3}) with {\it known functions $C^k$ and $B^a$}.

Consider consequently one-, two-, $\cdots$ $n$-dimensional algebras of operators (\ref{8.1})
which we write in the form
 \be \label{8.1}\tilde X=\mu D+N,\ \ \
N=C^{1}(u\p_u+v\p_v)+C^2u\p_v+B^1\p_{ u}+B^2\p_v .\ee

Let (\ref{8.1}) be a basis element of a one-dimensional algebra
$\cal A$ then commutators of $\tilde X$ with  $P_0$ and $P_a$ are
equal to a linear combination of $\tilde X$ and operators
(\ref{4.1}). This can happen in the following cases: \be
\label{8.2} \ba{l} C^{a}=\mu^{a}, \  B^a=\nu^a; \ \ \
C^{a}=e^{\lbd t} \mu^{a},
\ B^a=e^{\lbd t} \nu^a,\  \mu=0;\\
 C^{a}=0, \
B^a= e^{\lbd t+\om \cdot x} \nu^a, \  \mu=0\ea \ee where
$\mu^{ab}, \mu^a, \lbd $, and $\om=(\om_1,\om_2,\cdots, \om_m)$
are constants, $\om\cdot x=\om_\nu x_\nu$ and $\om\neq 0$.

To classify all non-equivalent symmetries (\ref{8.1}), (\ref{8.2}) we use
isomorphism of the related $N$
with $3
\times 3$ matrices of the following form \be \label{8.5} g= \left( \ba{ccc}
0   &  0       &  0\\
\nu^1 &  \mu^{1}  &  0\\
\nu^2 &  \mu^{2}  &  \mu^{1}  \ea \right). \ee

Equivalence transformations (\ref{x.2}) generate the following
transformation for matrix (\ref{8.5}) \be \label{8.6} g \to g'=U g
U^{-1}\ee where \be\label{8.51}  U=\left( \ba{ccc}
1   &  0       &  0\\
b^1 &  K^{1}  & 0\\
b^2 &  K^{2}  &  K^{1}  \ea \right),\qquad
U^{-1}=\frac1{K^1}\lo\ba{ccc}K^1&0&0\\-{b^1}&1&0\\{b^1K^2-b_2}&-
\frac{K^2}{K^1}&1\ea\ro. \ee

Up to transformations (\ref{8.6}) there exist six non-equivalent matrices $g$, i.e.,
\be\ba{l} \label{8.10}
g_1=\left( \ba{ccc}
0   &  0       &  0\\
0 &  1  &   0 \\
0 &  0  &  1 \ea \right),  \quad g_2=\left( \ba{ccc}
0   &  0       &  0\\
\lbd &  0  &   0 \\
1 &  0  &  0  \ea \right), \quad g_3=\left( \ba{ccc}
0   &  0       &  0\\
1 &  0  &   0 \\
0 &  0  &   0  \ea \right), \\\\
g_4=\lo\ba{ccc}0&0&0\\0&1&0\\0&1&1\ea\ro, \ \ \
g_5=\lo\ba{ccc}0&0&0\\0&0&0\\0&1&0\ea\ro, \ \ \
g_6=\lo\ba{ccc}0&0&0\\1&0&0\\0&1&0\ea\ro.\ea\ee

Let us denote
\[\hat g=g^{22}u\p_u+g^{33}v\p_v+g^{32}u\p_v+g^{21}\p_u+g^{31}\p_v\]
where $g^{ks}$ are elements of matrix $g$. Then in accordance with (\ref{8.2}) the
related symmetries (\ref{8.1}) have the form
\be\label{1}\ba{l}\tilde X=\mu D+\hat g\ \   \text{for}\
g=g_1, g_4, g_5,  g_6,\\
\tilde X=e^{\lbd t+\om \cdot x}\tilde g \ \ \text {for}\ g=g_1, g_2,\\
\tilde X=e^{\lbd t} \hat g \ \text{for any}\ g \ (\ref{8.10}).
\ea\ee

Formulae (\ref{8.10})-(\ref{1}) give the principal description of all
possible one-dimension algebras $\cal A$ which can be admitted by equation (\ref{1.2}).

To describe two-dimension algebras $\cal A$ we classify matrices
$g$ (\ref{8.5}) forming two-dimension Lie algebras. Up to
equivalence transformations (\ref{x.2}) there exist six such
algebras: \be\label{8.33}\ba{l} A_{2,1}=\{g_3,\tilde g_2\},\
A_{2,2}=\{g_1, g_5\},\ A_{2,3}=\{g_5,\tilde g_2\},\
A_{2,4}=\{g_6, \tilde g_2\},\ea\ee \be\label{8.34} A_{2,5}=\{g_1,
g_2\},\ A_{2,13}=\{g_1, g_3\}\ee where $\tilde g_2$ is matrix
$g_2$ (\ref{8.10}) with $\lbd=0$.

Algebras (\ref{8.33}) are Abelian while (\ref{8.34}) are
characterized by the following commutation relations: \be
\label{8.16} [e_1, e_2]= e_2 .\ee Two-dimension algebras $\cal A$
generated by (\ref{8.33}) and (\ref{8.34}) are spanned on the
following basis elements \be \label{8.17}\ba{l} < \mu D+ \hat
e_1+\nu  t \hat e_2, \hat e_2>,\ < \mu D+ \hat e_2+\nu  t \hat
e_1, \hat e_1>\\<\mu D-\hat e_1,
         \nu D-\hat e_2>,\  <F_1 \hat e_1+ G_1 \hat e_2,\ F_2\hat e_1+G_2
 \hat e_2> \ea\ee and
 \be\label{8.19} <\mu D- \hat e_1, \hat e_2>, \ < \mu D+ \hat e_1+\nu  t \hat e_2, \hat e_2>\ee respectively,
 where $\{F_1, G_1\}$ and $\{F_2,G_2\}$ are fundamental solutions of
       the following system \be \label{8.20}  F_t=\lbd F+\al G, \quad
       G_t = \sigma F+\gm G \ee with arbitrary parameters $\lbd, \al, \s,
       \gamma$. Arbitrary parameters $\mu$ and $\nu$ in particular can be equal to zero.

   In addition, there exist two dimension algebras $\cal A$ which are induced by
   one-dimension algebras of matrices $g$ (\ref{8.5}), namely
  \be\label{2}< F\hat g,
   G\hat g>,\ \ \ <\mu D+\lbd e^{\nu t +\om \cdot x}\hat g, e^{\nu t +\om \cdot x}\hat g>\ee
   with $F$ and $G$ satisfying (\ref{8.20}). Such algebras correspond to
   incompatible classifying equations (\ref{4.3}).

   Up to transformations  (\ref{8.6}) there exist four three-dimension algebras $A_{3,1}-A_{3,4}$
   of matrices (\ref{8.5}) and the only four-dimension algebra of such matrices which we denote as $A_4$:
\begin{center} { \bf Table 1. Three- and four-dimension algebras of matrices (\ref{8.5})}
\end{center}

  \begin{tabular}{|l|l|l|}
   \hline
    Algebra&Basis elements&Nonzero commutators\\
    \hline
    $A_{3,1}$&$
   e_1=g_1, \ e_2=g_3, \ e_3=\tilde g_2$& $[e_1,e_2]=e_2,\ \ [e_1,e_3]=e_3$\\
      $A_{3,2}$& $e_1=g_5, \ e_2=g_1, \ e_3=\tilde
      g_2$&$[e_2,e_3]=e_3$\\
      $A_{3,3}$&$ e_1=\tilde g_2, \ e_2=g_5, \ e_3=g_6$&$[e_2,e_3]=e_1$\\
      $A_{3,4}$&$ e_1=\tilde g_2, \ e_2=g_3, \ e_3=g_4$&$[e_1,e_2]=e_2,\
       [e_1,e_3]=e_2+e_3$ \\
       $A_4$&$\ba{l} e_1=g_1,\  e_2=g_3,  \ e_3=\tilde g_2,\\  e_4=g_5 \ea$&$\ba{l}[e_1,e_2]=e_2,\
       [e_1,e_3]=e_3,\\ \ [e_4,e_2]=e_3\ea$\\
       \hline
       \end{tabular}

Using commutation relations present in the table we come to the
following related three-dimension algebras $\cal A$ :
\[\ba{l}<\mu D-\hat e_1,\ \hat e_2,\  \hat e_3>, \
 <\hat e_1,\
\ F_1 \hat e_2+G_1 \hat e_3,\ F_2\hat e_2+G_2 \hat e_3>\ea\]
    with $e_a$ belonging to $A_{3,1}$;
\[<\mu D-2\hat e_1,\  \nu D-2\hat e_2, \hat e_3>\]
    with $e_a$ belonging to $A_{3,2}$;
\[\ba{l} <\mu D-2\hat e_2,\ \nu D-2\hat e_3,\ \hat
e_1>, \
<\hat e_1,\ D+2e_\al+2\nu t \hat e_1,\ \hat e_{\al'}>,\\
<e^{\nu t+\omega\cdot x}\hat e_1,\ e^{\nu t+\omega\cdot x}\hat
e_\al,\ \hat e_{\al'}> \ea \] where $\al,\ \al'=2,3, \al'\neq \al$
and $e_a$ belong to $A_{3,3}$;
    \[<\mu D-2\hat e_1, \ \hat e_2,\
\hat e_3>,\ <\hat e_1, \ e^{\nu t+\omega\cdot x}\hat e_2,\ \
e^{\nu t+\omega\cdot x}\hat e_3>\] with $e_a$ belonging to
$A_{3,4}$.

The four-dimensional algebra $A_4$ induces algebras $\cal A$ given
below:
\[\ba{l}
<\mu D-2\hat e_1,\ \nu D-2\hat e_4,\ \hat e_2,\ \hat e_3>,\ \
<e^{\nu t+\omega\cdot x}\hat e_1,\  e^{\nu t+\omega\cdot x}\hat
e_4,\ \hat e_2,\ \hat e_3>\ea.\]

Thus we had specified algebras of main symmetries which can be
admitted by equation (\ref{1.2}).

\section{Solution of classifying equations for the case of invertible diffusion matrix}

Applying  results of the previous section we can easily classify
main symmetries of equation (\ref{1.2}). Such classification
reduces to solving equations (\ref{4.3}) with their {\it known
coefficients} $C^{1}, C^2$ and $B^1, B^2$ which can be found
comparing (\ref{4.2}) with the found realizations of algebras
$\cal A$. To complete the group classification of equations
(\ref{1.2}) we will specify all cases when relations
(\ref{4.8})-(\ref{4.10}) are satisfied, i.e., when the main
symmetries can be extended.

Solving of equations (\ref{4.3}) with known $C^1, C^2$ and $B^1,
B^2$ is a rather routine procedure. We restrict ourselves to
presentation of an example of such solution in the following. Note
that asking for invariance of (\ref{1.2}) w.r.t. one-dimension
algebra $\cal A$ we fix the non-linearities $f^1$ and $f^2$ up to
arbitrary functions while an invariance w.r.t. a two-dimension
algebra $\cal A$ usually fixes these non-linearities up to
arbitrary parameters.

We will solve classifying equations (\ref{4.3}) up to equivalence
transformations $ U \to \tilde U=G(U,t,x)$, $t \to \tilde
t=T(U,t,x)$, $x \to \tilde x=X(U,t,x)$ and $ f \to \tilde
f=F(U,t,x,f)$ which keep the general form of equations (\ref{1.2})
but can change functions $f^1$ and $f^2$.

The group of equivalence transformations for equation (\ref{1.2})
can be found using the classical Lie approach and treating $f^1$
and $f^2$ as additional dependent variables. In accordance with
their definition, equivalence transformations include all symmetry
transformations (i.e., transformations generated by operators
(\ref{4.1}) and other symmetries  which will be found in the
following) and also transformations (\ref{x.2}). In addition, for
some particular non-linearities $f^1$ and $f^2$ there exist
additional equivalence transformations, whose list is given in
formulae (\ref{eqv}): \be\ba{lll}\label{eqv}
1.&&u\to\exp(\omega t)u, \ \ v\to \exp(\omega t)v,\\
2.&&u\to u+\omega t+\mu x^2,\ v\to v,\\
3.&&u\to u,\ v\to v+\rho t +\mu x^2,\\
4.&&u\to u+\rho t,\ v\to v\exp(\rho t),\\
5.&&u\to u,\ v\to v+\rho tu,\\
6.&&u\to \exp(\omega t) u,\ v\to v+\kappa tu+\rho \frac{t^2}{2},\\
7.&&u\to u,\ v\to v-\rho tu+\rho\lambda \frac{t^2}{2},\\
8.&&u\to \exp(\rho t)u,\ v\to \exp(\rho t)\lo v+\varepsilon\frac{t^2}{2}u\ro,\\
9.&&u\to u+\rho t,\ v\to v+\rho t u+\rho\frac{t^2}{2},\\
10.&&u\to\exp(\om t)u,\ v\to \exp(\om t)(v-\om t u),\\
11.&&\texttt{Transformations (\ref{eqv3}) valid for}\ a=0\
\text{only. }\\ \ea\ee where $\Phi(u)$ is an arbitrary function of
$u$, $F_1$ and $F_2$ are functions of $u$ which appear in the
classified equations.

In the following we specify additional equivalence transformations
(\ref{eqv}) admitted by some of equations (\ref{1.2}).

Let us present an example of solving of classifying equations.
Consider the first of algebras $\cal A$ given by relation
(\ref{8.17}) with $e_1,\ e_2$ belonging to algebra $A_{2,2}$
(\ref{8.33}). It includes two basis elements
\[X_1=\mu D-u\p_u-v\p_v,\ \ \ X_2=\nu D-u\p_v.\]

Comparing $X_1$ with $\tilde X$ (\ref{8.1}) we conclude that in
this case $C^1=1, C^2=B^1=B^2=0$ and so the classifying equations
(\ref{4.3}) are reduced to the following ones:
\[(\mu+1)f^a=(u\p_u+v\p_v)f^a,\ \ a=1,2\]
General solutions of this system have the form
    \be\label{n01}f^1=u^{\mu+1}F_1,\ \ f^2=v^{\mu+1}F_2\ee
where $F_1$ and $F_2$ are arbitrary functions of $\frac{u}{v}$.

Thus equation (\ref{1.2}) admits symmetry $X_1$ iff the related
non-linearities are of the form (\ref{n01}).

Asking for symmetry of equation (\ref{1.2}) w.r.t. transformations
generated by $X_2$ we come to the following classifying equations
(\ref{4.3}): \be \label{n02}\nu f^1=u\p_v f^1,\ \ \nu
f^2+f^1=u\p_v f^2.\ee

Substituting (\ref{n01}) into (\ref{n02}) we come to the equation
whose general solution for $\mu\neq 0$ is
\be\label{n03}\begin{array}{l} f^1=\lambda u^{\mu +1}e^{\nu
\frac{v}{u}},\ \ f^2=e^{\nu \frac{v}{u}}(\lambda v +\sigma
u)u^\mu.\ea\ee

In the special case $\mu=0$ the solution has the form
    \be\label{n04}
f^1=\lambda ue^{\nu \frac{v}{u}}+\omega u,\ \ f^2=e^{\nu
\frac{v}{u}}(\lambda v +\sigma u)+\omega v.\ee However, in this
case there exist the additional equivalence transformation
(\ref{eqv}) given in Item 1 with $\rho=\om$, which reduces
parameter $\om$ in (\ref{n04}) to zero. So without loss of
generality we can restrict ourselves to solutions (\ref{n03}) for
any $\mu$.

We see that equation (\ref{1.2}) admits the two-dimension algebra
of main symmetries  spanned on $X_1, X_2$ provided $f^1$ and $f^2$
have the form (\ref{n03}). This symmetry can be extended if
functions (\ref{n03}) and $f^2$ satisfy one of conditions
(\ref{4.8}), (\ref{4.10}) or both the conditions (\ref{4.8}),
(\ref{4.9}).

Equation (\ref{4.10}) is incompatible with (\ref{n03}). In order
equation (\ref{4.8}) be satisfied we have to impose the condition
$\mu=-a\nu$ on parameters $\mu, \nu$ and $a$. The related equation
(\ref{1.2}), (\ref{n03}) has the form
    \be\label{n06}\ba{l}
\displaystyle u_t-a\Delta u=f^1=\lambda u^{1-a\nu}e^{\nu
\frac{v}{u}},\\ v_ t-\Delta u-a\Delta  v=e^{\nu
\frac{v}{u}}(\lambda v +\sigma u)u^{-a\nu} \ea \ee and
 admits the Galilei
generators $G_\mu$ (\ref{2.6}). Finally, asking for equation
(\ref{4.9}) be satisfied we obtain one more condition
$\nu=-\frac4{a}m$ which guaranties invariance of equation
(\ref{n06}) w.r.t. the conformal generator $K$ of (\ref{2.6}).

The obtained classification results are presented in Table 2, Item
3.

In analogous way we solve classifying equations for other algebras
$\cal A$ and specify the cases when
the main symmetry can be extended.
\section{Classification results for equations (\ref{1.2})
with invertible diffusion matrix} The classification results are
given in the following  Tables 1-5 when we also indicate the
additional equivalence transformations (AET) (\ref{eqv}) which are
admitted by some particular equations (\ref{1.2}). The symbols
$D,\  \hat G_\nu,\  G_\nu$ and $K$ are used to denote operators
(\ref{2.6}). In addition, we denote
\[{\tilde
K}=K+\frac{1}{\lambda-1}\left(t\left(pu{\p_u}+(2-\lambda)v{\p_
v}\right) +u{\p_ v}\right).\]

To save a room we present in Table 2, Items 8-11 and Table 3, Items 1-3, the classification
results which are valid for equation (\ref{1.2}) with $a\neq0$ and $a=0$ as well. The completed analysis
of  symmetries and
the corresponding non-linearities for equation (\ref{1.2}) with $a=0$ is given
in Section 7.

In the following tables $F_1, F_2$ and $F$ are arbitrary functions whose arguments are specified
in the third column,   $\Psi(x)$ is an
arbitrary function of $x_1, \ x_2,\ \cdots,\ x_m$ and
$\psi_\nu(x)$ is a solution of the linear heat equation
$(\p_t-\Delta)\psi_\nu=\nu\psi_\nu$. In addition, we denote by $\Psi_\mu(x)$ a
solution of the Laplace equation $\Delta\Psi_\mu(x)=\mu\Psi_\mu(x)$.

Greek letters in the tables denote arbitrary parameters which can
take any (including zero) real values. The only exception is
parameter $\va$ in as much as without loss of generality we can
restrict ourselves to its values $\va=\pm1$.

\begin{center} { \bf Table 2. Non-linearities with arbitrary
functions and symmetries for equations (\ref{1.2}) }
\end{center}
  \begin{tabular}{|l|l|l|l|}
\hline \text{No} & \text{Nonlinear terms} &\begin{tabular}{c}
Argu- \\
ments\\ of \\
$F_1$, $F_2$\\
\end{tabular}
& {Symmetries}
\\
\hline 1. &$\ba{l} f^1=\nu u+F_1,\\ f_2=\nu u^2+F_1u+F_2,\ \ea$&$
2v-u^2 $&$ \psi_\nu\left( u{ {\p_v}}+{
{\p_u}}\right)  $\\
\hline 2.&$\ba{l} f^1=e^{\nu u}F_1,\\f^2=e^{\nu u}(F_2+F_1u),\
\ea$&$ 2v-u^2 $&$ \nu D-u{
{\p_v}}-{ {\p_u}} $\\\hline 3. &$\ba{l} f^1=F_1,\\ f^2=F_2+\nu v,\  \ea$&$ u $&$\ba{l}
\psi _\nu {\p_v}\ea
$\\
\hline 4.&$\ba{l} f^1=\alpha u+\mu ,\\ f_2=\nu v+F,\ \alpha \mu
=0\ea$&$ u $&$ \ba{l}\psi _\nu {\p_v},
\\e^{(\nu -\alpha )t}\left(u-\mu t\ro{\p_v} \ea $\\
\hline
5. &$\ba{l} f^1= u^2, \\
f^2=uv+\nu v+F\ea$&$ u $&$\ba{l} e^{\nu t} u {\p_v},\\
e^{\nu t}\left( {\p_v}+tu
{\p_v}\right)  \ea$\\
\hline
6. &$\ba{l} f^1= \left( u^2-1\right),\\
f^2=\left(  u+\nu \right) v+F\ea$&$ u $&$
\begin{array}{l}
e^{(\nu +1 )t}\left( u{ {\p_v}+  {\p_v}}\right) , \\
e^{(\nu -1 )t}\left( u{ {\p_v}-  {\p_v}}\right)
\end{array}
$\\
\hline
7. &$\ba{l} f^1= \left( u^2+1\right) ,\\
f_2=\left(  u+\nu \right) v+F\ea$&$ u $&$
\begin{array}{l}
e^{\nu t}\left( \cos  tu {\p_v}- \sin t {\p_v}\right),
 \\
e^{\nu t}\left( \sin tu {\p_v}+\cos t {\p_v}\right)
\end{array}$\\
\hline 8*. &$
\begin{array}{l}
f^1=e^{\nu v}F_1, \  f^2=e^{\nu v}F_2
\end{array}$
& $\ba{l}u\ea$ &$\nu D- {\p_v}$  \\
\hline 9*.&$
\begin{array}{l}
f^1=e^{\nu u}F_1, \\
f^2=e^{\nu u}F_2
\end{array}$
& $\ba{l}v\ea$ & $\nu D-{\p_u}$\\
\hline
 10*.&$\ba{l} f^1=\nu u+F_1,\\ f_2=-a\mu u+F_2 \ea$&$ v $&$
e^{( \nu +a\mu
)t}\Psi _\mu(x) { {\p_u}} $\\
\hline
    11*.&$\ba{l}f^1=u(F_1+\nu\ln u),\\f^2=v(F_2+\nu \ln u),\
\nu\neq0\ea$&$\frac{u}{v}$&$\ba{l}e^{\nu t}(u\p_u+v\p_v)\ea$\\
 \hline
\end{tabular}

\vspace{2mm}

The items marked by asterisks are valid for both cases $a\neq0$ and $a=0$.
If $a=0$ then in Item 8* without loss of generality $F_1=1$.

\newpage
\begin{center}
{ \bf Table 3. Non-linearities with arbitrary functions,
symmetries and AET  for equations (\ref{1.2}) }
\end{center}
  \begin{tabular}{|l|l|l|l|l|}
\hline \text{No} & \text{Nonlinear terms} &\begin{tabular}{c}
Argu- \\
ments\\ of \\
$F_1$, $F_2$\\
\end{tabular}
&

{Symmetries}& $ \ba{c}\text{AET} \\(\ref{eqv})\ea$
\\
    \hline
1*. &$
\begin{array}{l}
f^1=uF_1-\nu v, \\
f^2=\nu\frac{v }{u}(v-u) \\
+uF_2-vF_1,\ \nu\neq0
\end{array}$
& $ue^{\frac{v}{u}}$ &$
\begin{array}{l}
e^{\nu t}\left(u{ {\p_v}} - u{ {\p_u}}\right.\\\left.-v{ {\p_v}}\right)\\
\&\  \widehat G_\al\ \text{if}\ a=1
\end{array}$&$\ \ \ \ \ $
\\
\hline 2*. &$
\begin{array}{l}
f^1=u^{\nu +1}F_{1,} \\
f^2=u^\nu \left(F_2u- F_1v\right)
\end{array}$
& $ue^{\frac{v}{u}}$& $\begin{array}{l}\nu D+u{ {\p_v}}-u{
{\p_u}}\\-v{ {\p_v} }\ \& \ G_\al\\ \text{if}\ \nu=0,\ a=1
\end{array}
$&$\ \ \ \ \ $\\
\hline
 3*.&$\ba{l}f^1=u^{\mu+1}F_1,\\
f^2=u^{\mu+1}F_2\ea$&$\frac{v}{u}$&$\mu D-u\p_u-v\p_v$&$\ba{c} 1,\ \text{if}\\
  \mu=0\ea$\\
\hline
 4. &$
\begin{array}[t]{l}
f^1=e^{\nu {\frac{v}{u}}}F_1u, \\
f^2=e^{\nu {\frac{v}{u}}}\left( F_1v+F_2\right)
\end{array}$
&$u $& $\ba{l} \nu D-u{ {\p_v}}\ea$&$\ba{c} 5\ \text{if}\\\nu=0\ea$
\\
\hline 5. &$
\begin{array}{l}
f^1=u(F_1-\nu), \\
f^2=F_1v+F_2,\\\nu\neq0
\end{array}$
& $u$ & $\ba{l}e^{\nu t} u{ {\p_v}}\\   \&\ \psi_\nu\p_{v}\
\text{if}\  F_1=\nu \ea$&$\ba{c}3,\
\text{if}\\ F_1=0\ea$  \\
\hline
\end{tabular}

\vspace{3mm}


\begin{center}{\bf
Table 4. Non-linearities with arbitrary parameters and extendible
symmetries for equations (\ref{1.2})} with $a\neq0$
\end{center}
\begin{tabular}{|l|l|l|l|l|}
\hline No & \text{Nonlinear terms}
&$\ba{l}\text{Main}\\
\text{symmetries}\ea $&$
\begin{array}{l}
\text{Additional}\\\text{symmetries} \\
\end{array}$&$\ba{l}\text
{AET}\\(\ref{eqv})\ea$\\
\hline
  1. &$
\begin{array}{l}
f^1=\lambda u, \\
f^2=\sigma u^\mu\
\end{array}
$&$\ba{l} \psi _0 {\p_v},\\e^{ -\lambda t}u {\p_v} \ea$&$\ba{l}
e^{\lambda t}\left( u {\p_v}+\lambda { {\p_u}}\right) \\%
\text{ if }\mu =2\ea $&$\ba{c}3;\\ 5\ \text{if}\\  \lbd=0\ea$\\
\hline
 2. &$
\begin{array}{l}
f^1=\lambda e^{u}, \\
f^2=\sigma e^{u}\
\end{array}
$&$\ba{l} D-{{ {\p_u},}}\\\psi _0 {\p_v}\ea $&$  u{ {\p_v}}\
\text{ if }\lambda =0
$&$\ba{c}3;\\ 5\ \text{if}\\  \lbd=0\ea$\\
\hline
 3. &$
\begin{array}{l}
f^1=\lambda u^{\nu +1}e^{\mu \frac{v}{u}},\\
f^2=e^{\mu \frac{v}{u}}(\lambda v +\sigma u)u^\nu\ea$&$\ba{l} \mu
D-u{ {\p_v},}\\\nu D-u{ {\p_u}}\\-v{ {\p_v}}\ea$&$
\begin{array}{l}
 G_\alpha\ \text{ if }\nu =a\mu\\
\text{\&}\ K\ \text{if}\ \nu =\frac 4m
\end{array}
$&$\ba{c}1,\ \text{if}\\ \nu=0;\\5,\ \text{if}\\\mu=0 \ea$ \\
 \hline
 4.&$\ba{l}f^1=\lbd u^{\mu+1},\\ f^2=\s u^{\mu+1},\\
\ \lbd\s=0\ea$&$\ba{l} \mu D- u{\p_u}\\-v{\p_ v},\\\psi_0 {\p_
v}\ea$&$\ba{l}  u{\p_ v}\ \text{if}\ \lbd=0\ea$&$\ba{c}3; \\
5\
\text{if}\\ \lbd=0 \ea $\\
\hline
\end{tabular}

\newpage

\begin{center}{\bf
Table 5. Non-linearities with arbitrary parameters and
non-extendible symmetries for equations (\ref{1.2})} with $a\neq0$
\end{center}

\begin{tabular}{|l|l|l|l|}
\hline &$ \text{Nonlinear terms} $ &$ \text{Symmetries} $&
$\ba{l}\text{AET}\\(\ref{eqv}) \ea$\\
\hline 1. &$\ba{l}f^1=\lambda v^{\nu +1},\\f^2=\mu v^{\nu +1}\ea
$&$\ba{l} \nu D-u
 {\p_u}-v {\p_v},\\
\Psi_0
(x) {\p_u} \ea$&$\quad\ 2$\\
\hline 2. &$\ba{l} f^1=\lambda e^{v},\ f^2=\sigma e^{v} \ea$&$
\ba{l}D- {\p_v},\ \Psi_0 (x) {\p_u}\ea$&$\quad\ 2$\\
\hline 3. &$\ba{l} f^1=\lambda e^{u},\ f^2=\sigma ue^{u}\\\
\ea
$&$\ba{l} D-{{%
 {\p_u}-2}}u {\p_v},\
\psi _0 {\p_v}\ea $&$\ba{c} 3;\\ 5\ \text{if}\\  \lbd=0\ea$\\
\hline 4.&$
\begin{array}{l}
f^1=\nu e^{\lambda \left( 2v-u^2\right) }, \\
 f^2=\left( \nu u+\mu \right) e^{\lambda \left( 2v-u^2\right)
}
\end{array}
$&$\ba{l} \lambda D- {\p_v},\
{\p_u}+u{ {\p_v}}\ea $&$\quad\ 9$\\
\hline 5.&$ \ba{l} f^1=\mu \ln v,\\ f^2=\nu \ln v
\ea$&$
\begin{array}{l}
\Psi_0 (x){\p_u}
,\  D+u{ {\p_u}}+v{ {\p_v}}\\
{+}\left( (\mu- \nu a)t-\frac \nu {2m}mx^2\right){\p_u}
\end{array}
$&$\quad\ 2$\\
\hline 6.&$\ba{l} f^1=\lambda\\ f^2=\varepsilon\ln u ,\ \ea$&$
\begin{array}{l}
 D+u{ {\p_u}}+v{ {\p_v}}{+\varepsilon t
{\p_v}},\\ \psi _0 {\p_v},\ \left( u-\lambda t\right) { {\p_v}}
\end{array}
$&$\ba{c}3,\ 7;\\\&\ 5\ \texttt{if}\\\lbd=0\ea$\\
\hline
 7.&$ \ba{l}f^1=\lambda u^{\nu +1},\\ f^2= \lambda u^{\nu
+1}\ln u,\\   \nu\neq-1 \ea $&$ \ba{l}\nu D-u{ {\p_u}} -v {\p_v}-u{ {\p_v}},\\
\psi _0 {\p_v} \ea$&$\ba{c} 3\ea$\\
\hline 8.&$\begin{array}{l}
f^1= \varepsilon(2 v- u^2), \\
f^2=(\mu+ \va u) \left( 2v-
u^2\right)\\-\va\frac{\mu^2}{2}u,\ \ \mu\neq0\end{array} $&$\ba{l}X_1=e^{\mu t}
\lo 2
{\p_u}+2 u{\p_ v}\right.\\\left.+\mu\va {\p_ v}\ro,\ \
2tX_1+\va e^{\mu t}{\p_ v}\ea$&$$ \\
\hline 9.&$\begin{array}{l}
f^1= \varepsilon(2 v- u^2), \\
f^2=(\mu+\va u) \left( \varepsilon v-
u^2\right)\\+\varepsilon\frac{1- \mu^2}{2}u\end{array} $&$\ba{l}X^{\pm}=e^{\mu\pm 1}\lo
2{\p_u}+2 u{\p_ v}\right.\\\left.+\varepsilon(\mu\pm 1)
 {\p_ v}\ro\ea$
&$\ba{c}9,\text{if}\\\mu^2=1\ea$\\
\hline 10.&$\begin{array}{l}
f^1= \varepsilon (2v-u^2), \\
f^2=(\mu+\varepsilon u) \left(
2v-u^2\right)\\-\varepsilon \frac{1+\mu^2}{2}u \end{array} $&$\ba{l}e^{\mu t}\lo
2\cos t\lo{\p_u}+u{\p_
v}\ro\right.\\
\left.+\varepsilon (2\mu \cos t-\sin t){\p_
v}\ro,\\
e^{\mu t}\lo 2\sin t\lo{\p_u}+ u{\p_
v}\ro\right.\\
\left.+\varepsilon (2\mu \sin t+\cos t){\p_ v}\ro
\ea$&\\
\hline
\end{tabular}

\vspace{3mm}

In the following table we present symmetries of a special subclass of
equations (\ref{1.2}) whose r.h.s. is given in the table title.
The related non-linearities had appeared in the classification
procedure of reaction-diffusion equations with the unit diffusion
matrix \cite{N2} where they corresponded to a very rich spectrum of
symmetries.

In the case of triangular diffusion matrices studied in the
present paper the classification results present in Table 5
include thirteen non-equivalent types of equations. Among them
there are five equations invariant w.r.t. two dimensional algebras
of main symmetries, and four equations admitting symmetries
$G_\nu$ i.e., being invariant w.r.t. the Galilei group.

In Table 5 the following notations are used: $\delta=\frac
14(\mu-\nu)^2+ \lambda\sigma, \  \om_0=\frac12(\mu+\nu), \ \ \
\om_\pm=\om_0\pm 1
$
where $\mu,\ \nu,\ \lbd$ and $\s$ are parameter used in definition
of the related non-linearities $f^1,\ f^2$.
\begin{center}
\textbf{Table 6.  Symmetries of equations (\ref{1.2}) with
non-linearities $f^1=\lambda v+\mu u \ln u$, $ f^2=\lambda \frac
{v^2}{u}+(\s u+\mu v)\ln u+\nu v$ and $a\neq0$}
\end{center}
\begin{tabular}{|l|l|l|l|}
\hline
No&Conditions &Main symmetries&Additional\\
&for coefficients && symmetries \\
\hline $1$&$\ba{c}\lambda=0,\\ \mu\neq \nu\ea$&$\ba{l}e^{\nu
t}u{\p_ v}, \ea$&$\ba{l} \psi_\nu\p_v\ \text {if}\ \mu= 0,\\
\&\ G_a \texttt{if}\ \nu=a\s
\ea$\\
\cline{4-4} &&$\ba{l}e^{\mu t}\lo (\mu-\nu)R{\p_ R}+
\s u{\p_ v}\ro\ea$&$\ba{l}\hat G_a, \text{if}\ \mu \neq 0,\\
 a\s=\nu-\mu, \ea$
\\
    \hline
$2$&$\ba{l}\s=0,\\\mu\lambda\neq 0, \ea$&$e^{\nu t}\lo \lambda R{\p_
R}
+(\mu-\nu)u{\p_ v}\ro$&$\ba{l} G_a\, \text{if}\ \nu=0,\\ a\mu=-\lambda\ea$\\
\cline{4-4} &$\mu\neq \nu$&$ e^{\mu t}R{\p_ R}$&
$\ba{l}\hat G_a,\ \text{if}\ a(\nu-\mu)=\lambda\ea$\\
    \hline
    $3$ & $
\begin{array}{l}
\delta =0 \\
\end{array}
$ & $
\begin{array}{l}
X_4=e^{\omega_0 t}\left( 2\lambda  R{\p_ R}\right.\\\left. +(\nu
-\mu )u {\p_v}\right) ,
\end{array}
$ &$\begin{array}{l} G_a,\ \text{if}\ \omega_0 =0,\\ a\nu
=-\lambda\neq 0
 \ea$\\
\cline{4-4} &&$2e^{\omega_0 t}u
{\p_v}+tX_4$&$\ba{l}\widehat{G}_a,\ \text{if }\ \omega_0 \neq
0,\\2\lambda =a(\mu
-\nu )\neq 0\ea$\\
    \hline
    $4$&$\ba{l}\lambda\neq 0,\\ \delta=1\ea $ &$
\ba{l}e^{\omega_+ t}\lo \lambda R{\p_ R}+(\omega_+-\mu)u{\p_
v}\ro,\ea $&$\ba{l} G_a,\ \text{if} \ a\mu=\lambda\\
\omega_+\omega_-=0
\ea$\\
\cline{4-4} &&$e^{\omega_- t}\lo \lambda R{\p_
R}+(\omega_--\mu)u{\p_
v}\ro$&$\ba{l}\hat G_\al, \ \text{if}\ \omega_+\neq 0, \\
\lambda =-a(\omega_+-\mu)
 \ea$\\
\hline $5$ & $
\begin{array}{l}\lbd\neq0,\\
\delta =-1
\end{array}
$ &$
\begin{array}{l}
e^{\omega_0 t}[2\lambda \cos  tR{\p_ R}\\+((\nu -\mu )\cos
 t
-2\sin t)u {\p_v}], \\
e^{\omega_0 t}[2\lambda \sin tR{\p_ R}\\ +((\nu -\mu )\sin
t +2\cos  t)u {\p_v}]\ea$&none\\
\hline
\end{tabular}

\vspace{3mm}

If $\lbd=\mu=0$ or $\lbd=\nu=0$ then the related equation (\ref{1.2})
admits additional equivalence transformations 10 or 5 from the list
(\ref{eqv}) correspondingly.

 We see that there exist a number of non-equivalent
systems (\ref{1.2}) with non-degenerate diffusion matrix. Rather
surprisingly the number of such equation with nilpotent diffusion
matrix (which are classified in the following section) appears to
be  even more large.

\section{Group classification of reaction-diffusion equations with nilpotent
 diffusion matrix}
\subsection{Equations with first derivatives in $x$}

Consider now equations (\ref{1.3}), (\ref{1.31}) and specify their
Lie symmetries. In this subsection we restrict ourselves to the
case $p\neq0$ when generators of admitted Lie group have the
general form (\ref{n001}) while the related classifying equations
are given by formula (\ref{2.12}). Moreover without loss of
generality we put $p=1$.

We solve the classifying equations using the technique developed
in Sections 5 and 6. The general analysis of admissible algebras
$\cal A$ can be carried out in complete analogy with Section 4.
Moreover, the results present in Section 4 can be extended to the
case of equations (\ref{1.3}), (\ref{1.31}) provided we make a
formal change $D\to \tilde D=3t\p_t+2x_\nu \p_\nu-v\p_v$ in all
formulae where the operator $D$ appeared, and exclude all algebras
$\cal A$ where matrices $g_4, g_5$ and $g_6$ (\ref{8.10}) appear.
Of course it is necessary to take into account that in contrast
with $D$ operator $\tilde D$ does not commute with $\p_v$. As a
result we come to the following one-dimension algebras \be
\label{7.1}
\ba{l} \tilde X_1^{(1)}=\mu \tilde D- u\p_u-v\p_v, \\
\tilde X_1^{(2)}= \tilde D-\nu \p_u, \quad \tilde
X_2^{(\nu)}=e^{\nu t} \left( u\p_u+v\p_v\right),
\\
\tilde X_1^{(3)}= \tilde D+ u\p_u+v\p_v+\nu\p_v,\
 \
\tilde X_3^{(3)}=e^{\sigma_3 t+\rho_3 x}
\left(\p_u+\p_v\right),\\
\tilde X_3^{(1)}=e^{\sigma_1 t+\rho_1\cdot  x} \p_u,\ \  \tilde
X_3^{(2)}=e^{\sigma_2 t+\rho_2\cdot  x} \p_v \ea \ee and
two-dimension algebras \be \label{7.2} \ba{l} \tilde A_1=<\tilde
D, \tilde X_2^{(0)}>, \
\tilde A_2=<\tilde X^{(2)}_1, X^{(3)}_3>, \
\tilde A_3=<\tilde X^{(3)}_1, \tilde X^{(1)}_3>,\\ \tilde
A_4=<\tilde X^{(1)}_1, \tilde X^{(1)}_3>, \ \tilde A_6=<\tilde D+4( u\p_u+v\p_v)
+t \p_v, \quad X^{(2)}_3>,\\ 
 \tilde A_5=<\tilde X^{(1)}_1, \tilde X^{(2)}_3>,\ 

\tilde A_7=<\tilde D+3(u\p_u+v\p_v) + t \p_u, \quad X^{(1)}_3>.
\ea \ee

In this way the problem of group classification of the equations
with the first order derivative terms reduces to solving the
classifying equations (\ref{2.12}) which their known coefficients
$B^1, B^2, F$ and specifying the case when these equations have
non-trivial solutions. These coefficients are easily identified
comparing (\ref{7.1}), (\ref{7.2}) with (\ref{2.11}). For example,
for symmetry $\tilde X_1^{(1)}$ we have $F=1, B^1=B^2=0$, for
$\tilde X_1^{(2)}$ the values of these coefficients are $B^1=\nu,
B^2=F=0, \mu=1$, etc.  Solving the related classifying equations
(\ref{2.12}) we easily find the related non-linearities $f^1$,
$f^2$ which are given in Table 7.

In accordance with the results present in Table 7 equations
(\ref{1.3}), (\ref{1.31}) with $p\neq 0$ can admit neither Galilei
nor conformal symmetry transformations. This result follows
directly from formulae (\ref{n001}), (\ref{2.11}).
equations.

In six cases enumerated in the table the corresponding equations
(\ref{1.3}), (\ref{1.31}) admit infinite dimension symmetry
algebras whose generators are defined up to arbitrary functions 
or arbitrary solutions of linear equations,
see Items 5-7, 9-14 here. However, these infinite symmetries
generate the same classifying equations as one dimensional
algebras $X_3^{1}-X_3^{(3)}$ and two dimensional algebras  $\tilde
A_3, \tilde A_4, \tilde A_7$.

Finally we note that in addition to their symmetries,  equations
(\ref{1.3}), (\ref{1.31}) with $p\neq0$ admit equivalence
transformations ({\ref{x.2}) with $K^2=0$. These transformations
which  change functions $f^1, f^2$ and also the value of parameter
$p$ where used to simplify generators (\ref{7.1}), (\ref{7.2}) and
the corresponding equations (\ref{1.3}), (\ref{1.31}). Equations
(\ref{1.3}), (\ref{1.31}) with the non-linearities given in Items
3 (when $\nu=0$) and 8 of Table 6 admit additional equivalence
transformation $u \to e^{\sigma t} u,\ \ \ v\to e^{\sigma t}v$.
\newpage
\begin{center}{\bf
Table 7. Non-linearities and symmetries for equations (\ref{1.3}),
(\ref{1.31}) with $p=1$}
\end{center}
\begin{tabular}{|l|l|l|l|}
\hline & $ \mbox{Non-linearities} $ & $\mbox{Arguments} $ &
$\mbox{Symmetries} $\\
  &            & $\mbox{of}\ F_1\, F_2$  &     \\
\hline 1. & $f^1=u^{1+3\mu}_1 F_1$, & $v u^{-\mu-1}$ & $\ba{l}\\2\mu
\tilde D-u{\p_u}-v{\p_v}\ea$\\
  &  $f^2=u^{ 1+4\mu } F_2$ &         &               \\
\hline 2. & $f^1=v^3 F_1 $, & $u-\nu \ln v$ & $\ba{l}\\
\tilde D-\nu {\p_u}\ea$\\
  &  $f^2=u^4_2 F_2$  &         &               \\
\hline 3. & $ f^1=u(F_1+ \nu \ln u),$ & $\frac{v}{u}$ &
$\ba{l}\\e^{\nu t}\left( u{\p_u}+ v{\p_v}\right)\ea$\\
  & $ f^2=v(F_2+\nu \ln u)$ &   & \\
\hline 4. & $f^1=u^{-2}_1F_1,$ & $v-\nu \ln u$ & $\ba{l}\\\tilde D+u
{\p_u}+v\p_v+ \nu {\p_v}\ea$ \\
  & $f^2=u^{-3}_1F_2$ &   & \\
\hline 5. & $f^1=\lbd u +F_1,$  & $v$  & $\ba{l}\\e^{\lbd t}\Psi_\mu(x)
{\p_u}\ea$ \\
  & $f^2=-\mu u+F_2$  &   & \\
\hline 6. &  $f^1=\nu v+F_1,$  &  $u$  &  $\ba{l}\\e^{\lbd t-\nu
x_m}\Psi(\tilde x)
{\p_v}\ea$\\
   &  $ f^2=\lbd v+F_2 $    &  &  \\
\hline 7. &  $f^1=\lbd u+F_1$  &  $u-v$  & $\ba{l}\\
e^{\lbd t}\ e^\frac{x_m+t}{2}\Psi_\mu(\tilde x, x_m+t)
\left({\p_u}+{\p_v}\right),\ea$ \\
 & $f^2=\s v+F_2 $  &   &$\mu=\lbd-\s+\frac14$  \\
\hline 8. & $\ba[t]{l}f^1=\al u^{-2}_1 u^{3}_2, \\f^2=\nu u^{-3}_1
u^4_2\ea$ && $\ba{l}\\\tilde D, \quad\ u{\p_u}+v{\p_v}\ea$\\
\hline 9. & $\ba[t]{l}f^1=\al e^{3u},\\ f^2=\nu e^{4u} \ea$ &&
$\ba{l}\\\tilde D-{\p_u},\ \Psi(\tilde x){\p_v}\ea$ \\
\hline 10. & $\ba[t]{l}f^1=\al e^{-2 v},\\ f^2=\nu e^{-3 v}\ea$ &&
$\ba{l}\\\tilde D+u {\p_u}+v\p_v+{\p_v},\ \Psi_0(x)
{\p_u}\ea$ \\
 \hline 11. & $\ba[t]{l}f^1=\al u^{3\mu+1}, \\ f^2=\nu u^{4\mu+1}\ea$ &&
$\ba{l}\\\mu \tilde D-u{\p_u}-v {\p_v}, \ \ \Psi(\tilde x){\p_v}\ea$\\
\hline 12. & $\ba[t]{l}f^1=\al v^{2\nu+1}, \\ f^2=\nu
v^{3\nu+1}\ea$ && $\ba{l}\\\nu \tilde D-
 u {\p_u}-v {\p_v},\ \Psi_0 (x)
{\p_u}\ea$ \\
 \hline 13. & $\ba{l}f^1=\al u^{\frac{1}{4}}, \\ f^2=\nu \ln u\ea$
&& $\ba{l}\frac14\tilde D+v{\p_v}+u{\p_u}+\nu
t{\p_v}, \ \Psi(\tilde x){\p_v}\ea$\\
\hline 14. & $\ba{l}f^1=\nu \ln v, \\ f^2=\frac{\al}{\sqrt v}\ea$
&& $\ba{l}\frac13\tilde D+u{\p_u}+ v {\p_v}+\nu t
{\p_u},\ \Psi_0(x){\p_u}\ea$\\

\hline
\end{tabular}\\

Here $\Psi_\mu(x)$ and $\Psi(\tilde x, x_m+t)$ are arbitrary
solutions of the Laplace equation
 $\Delta \Psi_\mu=
\mu\Psi_\mu$ in $m$-dimensional space, $\tilde\Psi_\mu(\tilde x)$
is a solution of the Laplace equation in $m-1$-dimensional space,
$\tilde x=(x_1,x_2,\cdots, x_{m-1})$,  $\mu, \ \nu$ and $\lbd$ are
arbitrary parameters satisfying $\nu \lbd \not=0$. Finally, we
denote $\tilde D=3t\p_t+2x_\nu \p_\nu-v\p_v$.

\subsection{Equations (\ref{1.2}) with $a=0$}

The procedure of classification of equations (\ref{1.3}),
(\ref{1.31}) with $p=0$  (or equations (\ref{1.2}) with $a=0$)
appears to be more complicated then in the case of $p$ non-zero.
The general form of symmetry admitted by this equation is given by
equation (\ref{new}) while the classifying equations take the form
(\ref{last}).

A specific property of symmetries (\ref{new}) is that in contrast with operators (\ref{2.4}) and
(\ref{n001}) they can generate non-linear transformations for dependent variables
since the coefficient $B^3$ can be a function of $u$. In addition, the classifying equations
critically depend on the
number $m$ of independent variables $x_\nu$ and are qualitatively
different for the cases $m=1, m=2$ and $m>2$. Nevertheless, these
equations can be effectively solved with using the analysis of low
dimension symmetry algebras present in Section 4.

One more specific point in the classification of equations
(\ref{1.2}) with $a=0$ is that they admit powerful equivalence
relations
    \be\label{eqv2}u\to u, \ v\to v+\Phi(u)\ee
and
     \be\label{eqv3}u\to u, \ v\to v+\hat\Phi(u,t,x)\ee
 which did not appear
in our analysis presented in the previous sections.

Equivalence transformation (\ref{eqv2}) (with $\Phi(u)$ being an
arbitrary function of $u$) are admitted by any equation
(\ref{1.2}) with $a=0$. Transformations (\ref{eqv3}) are valid for
the cases when $f^1$ does not depend on $v$ and in the same time
$f^2$ is either linear in $v$ or does not depend on this variable.
Moreover, the related functions $\hat\Phi(u,t,x)$  should satisfy
the following system of equations
    \be\label{eqv1} \ba{l}f^2_v\hat\Phi_t-\hat\Phi_{tt}-f^1\hat\Phi_{tu}=0,\\
f^2_v\hat\Phi_{x_\nu}-\hat\Phi_{tx_{\nu}}-f^1\hat\Phi_{ux_{\nu}}=0\ea\ee

Thus the group classification of equation (\ref{1.2}) with $a=0$
is reduced to solving the classifying equations (\ref{last}) with
using the algorithm presented in Section 3. We will not reproduce
here  cumbersome and rather routine calculations which are needed
to classify equations (\ref{1.2}) with $a=0$ which can
be carried out in the same way as in Sections 2-6 and taking into
account the specific points mentioned in the above. The
classification results are presented below in Tables 8-10 and also
in Table 2, Items 8-11 and  Table 3, Items 1-3. The related items
are marked by asterisks. The additional equivalence
transformations are specified below the Tables 8,9 and 10.

In Tables 8-10 the symbol $W$ denotes a function of $t, x$ and $u$
which solve the following equation:
\[f^2_v-W_t-W_uf^1=0.\]
\begin{center}{\bf
Table 8. Non-linearities with arbitrary functions for equations
(\ref{1.2}) with nilpotent diffusion matrix}
\end{center}
\begin{tabular}{|l|l|l|l|}
  \hline
  No&Nonlinear terms&\begin{tabular}{l}
Argu- \\
ments\\ of \ $F_\al$
\end{tabular}&Symmetries\\
  \hline
1.&$\ba{l}f^1=F_1u^{\mu-\nu},\
f^2=F_2u^{\mu}\ea$&$\frac{u^{\nu+1}}{v}$&$\ba{l}Q_1=(\mu-1) D-\nu
t{\p_ t} -u{\p_u}\\-(\nu+1) v{\p_ v}\ \&\
(m-2)x^2\p_{x_a}\\-x_aQ_1\
\texttt{if}\ \nu(m-2)=4,\\ \mu(m-2)=m+2,\ m\neq 2\ea$\\
    \hline
           2.&$\ba{l}f^1=F_1uv^{\mu-1},\  f^2=F_2v^\mu, \\ F_2\neq0\ea$
&$u$&$\ba{l}\mu D-t{\p_ t}-v{\p_ v}\  \& \\ e^W\p_v\ \texttt{if}\
\mu= 1\  \&\\ H^a{\p_ {x_a}} - { H^b_{x_b}} v{\p_ v}\ \texttt{if}\
m=2\ea$\\
    \hline
3.&$f^1=F_1v^{-1},\ f^2=F_2+\nu v$&$u$&$\ba{l}e^{\nu t}\lo{\p_ t}+
\nu v{\p_v}\ro\  \&\ e^W\p_v  \ \texttt{if}\\ F_1=0\ea$\\
    \hline
            4.&$f^1=F_1v^{\mu-1},\ f^2=F_2v^{\mu}$&$ve^u$&$\mu D-t{\p_ t}
-v{\p_ v}+{\p_u}$\\
     \hline
        5.&$\ba{l}f^1=\frac{F_1}{v}+\nu ,\ f^2=F_2+\nu v\ea$&$ve^u
$&$e
^{\nu t}\lo {\p_ t}+\nu v{\p_ v}-\nu{\p_u}\ro$\\
\hline 6.&$\ba{l}f^1=0,\ f^2=F_2\ea$&$v$&$\ba{l}\Psi_0(x){\p_u},
    \
    x_a{\p_{ x_a}}+
2u{\p_u}\ea$\\
    \hline
        7.&$f^1=F_1,\ f^2=0$&$u$&$e^W{\p_ v},\ \
    x_a{\p_ {x_a}}-
2v{\p_ v}$\\
    \hline
        8.&$\ba{l}f^1=\frac{\nu}{\mu-1}u+F_1u^{2-\mu},
\\f^2=\frac{\mu\nu}{\mu-1}v+F_2u,\ \mu\neq1\ea$&
$vu^{-\mu}$&$\ba{l}e^{\nu t}\lo(1-\mu)t{\p_ t}-\nu u{\p_u}\right.\\\left.-
\nu\mu v{\p_ v}\ro\ea$\\
    \hline
    9.&$\ba{l}f^1=uF_1,\ m=1\\f^2=vF_2+u\ea$&$vu^3$&
$\ba{l}Q_2=\cos (2x)\lo
u{\p_u}-3v{\p_ v}\ro\\+\sin (2x)x{\p_ x},\  \ Q_3=(Q_2)_x\ea$\\
    \hline
    10.&$\ba{l}f^1=uF_1,\ m=1\\f^2=vF_2-u\ea$&$vu^3$&
$\ba{l}Q_4=e^{2x}\lo {\p_ x}+
u{\p_u}-3v{\p_ v}\ro,\\
Q_5=e^{-2x}\lo{\p_ x}-
u{\p_u}+3v{\p_ v}\ro\ea$\\
    \hline
        11.&$\ba{l}f^1=F_1, f^2=vF_2, \ m=2\ea$&$ve^u
$&$\ba{l}H^a{\p_{ x_a}}
- { H^b_{x_b}}\lo v{\p_ v}-{\p_u}\ro\ea$\\
\hline 12.&$\ba[t]{l}f^1=\nu e^\frac{ v}{u},\ f^2=e^\frac{
v}{u}F\ea$&$\ba{l}u\ea$&$\ba{l}\\D-u\p_v\ea$\\\hline
 13.&$f^1=F_1,\ F_2=vF_2+F_3$&$u$&$e^W\p_v$\\
\hline
\end{tabular}

\vspace{2mm}

For the non-linearities enumerated in Items 2 (when $\mu=1$), 3
(when $F_1=0$), 4 and 8 of Table 8 the related equation
(\ref{1.2}) with $a=0$ admits additional equivalence
transformations (\ref{eqv3}). In addition, transformations
(\ref{x.2}) and (\ref{eqv2}) and some equivalence transformations
from the list (\ref{eqv}) are admissible, namely, transformations
(2) for the non-linearities given in Item 1 (when $\nu=-1, \mu=0$)
and in Item 6,  transformations (1) and (3) for the
non-linearities from Item I (when $\nu=1, \mu=0$) and Item 7
respectively.


\begin{center}{\bf
Table 9. Non-linearities with arbitrary parameters and extendible
symmetries for equations (\ref{1.2}) with nilpotent  diffusion
matrix}
\end{center}
\begin{tabular}{|l|l|l|l|c|}
  \hline
    No & Non-linearities  & $\ba{l}\text{Main}\\ \text{symmetries}\ea$&$\ba{c}
\text{Additional}
  \\ \text{symmetries}\ea$&$\ba{c}\text{AET}\\(\ref{eqv})\ea$ \\
  \hline
    1. & $\begin{array}{l}
f^1=\lambda u^{\nu +1}v^\mu , \\
f^2=\sigma u^\nu v^{\mu +1}
\end{array}$ &  $\ba{l}(\mu+\nu)t{\p_ t}\\-
(\mu+1)u {\p_u}\\+(\nu-1)v\p_v\ea$&$\ba{l}x_aQ_6-2\kappa x^2{\p_{
x_a}}\ \text{if}\\ \kappa(m+2)
=\nu,\\ \kappa(2-m)=\mu\ea$&\\
    \cline{4-5}
    &&$\ba{l}Q_6=2\mu u {\p_u}\\+(\mu+\nu)x_a{\p_{ x_a}}\\-2\nu v{\p_
v}\\\ea$&$\ba{l} e^W\p_v\ \text{if}\ \lambda=0,\\ \mu=-1\
\text{\&}\ 2x^2{\p_{ x_a}}\\  -(m-2)x_aQ_6\ \text{if}\\
\nu=\frac{m+2}{m-2},\ m\neq 2\ea$&$\ba{c}11,
\\3,5\  \ea$
 \\
    \cline{4-5}
    &&&$\ba{l}\Psi_0(x){\p_u},\ \text{if}\\ \s=0,\nu=-1,\ \text{\&}\\  x_aQ_6
    +\frac{2}{m+2}x^2{\p_{ x_a}}\\
\text{if}\ \mu=\frac{m-2}{m+2},\ m\neq 2\ea$&$\ba{c}11,
\\2\  \ea$\\
    \cline{4-5}
    &&&$\ba{l}\Psi_0(x){\p_u}\\ \text{if}\ \lambda=\nu=0\ea$&$\ba{c}11,\\
2\  \ea$
     \\
     \cline{4-5}
     &&$\ba{c}\ea$&$\ba{l}e^W \p_v \
\texttt{if}\ \mu=0\  \&\\ H^a{\p_{ x_a}} - H^a_{x_a} v{\p_
v}\\\text{if} \  m=2\ea$&$\ba{c}11\ \&\\5\  \texttt{if}\\ \lbd=\s\ea$\\
     \hline 2. & $\ba{l} f^1=\lambda u^{\nu+1}v^{-1},\\f^2=\s u^\nu+\va
v,\ea $&
  $\ba{l}e^{\va t}\lo{\p_ t}+\va v {\p_
v}\ro,\\{Q'}_6=Q_6|_{\mu=-1}\ea$&$\ba{l}x_a{
Q'}_6-\frac{2}{m-2}x^2
{\p_{ x_a}}\\\text{if}\ \nu=\frac{m+2}{m-2},\ m\neq 2\ea$&\\
   \cline{4-5}
&$\lbd\neq0$&&$\ba{l}\Psi_0(x)\p_u\  \texttt{if}\\ \s=0,\
\nu=-1\ea$&$\ba{c}
2\  \ea$\\
    \cline{4-5}
&&&$\ba{l}v\p_v+u\p_u\\ \texttt{if}\ \s=0,\ \nu=1\ea$&$\ba{c}1\
 \ea$\\ \hline
 3.&$
\begin{array}{l}
f^1=\lambda e^{\nu u}, \\
f^2=\sigma e^{(\nu +1)u},
\end{array}
$&$\ba{l} (\nu+1) D-v {\p_v}\\-t{\p_ t}-{ {\p_u}},\ e^W\p_v\ea$&$
\begin{array}{l}
\nu\lo v{ {\p_v}}+t{\p_ t}\ro- {\p_u}\\ \text{ if }\sigma =0
\ea$&$\ba{c}11,\\3\  \ea$
\\
    \cline{4-5} &$\ba{l}\lbd\s=0\\\\\ea$&&$\begin{array}{l}
 v{ {\p_v}}+t{\p_ t}\\
 \text{ if }\ \lbd =0 \ea$ &$\ba{c}11,\\3,5\  \ea$\\
    \hline
        4. &$
\begin{array}{l}
f^1=\lambda e^{(\nu+1) v}, \\
f^2=\sigma e^{\nu v}
\end{array}
$&$ \ba{l}(\nu-1)D- u {\p_u}\\+t {\partial_ t}-
 {\p_v}, \\  \Psi_0(x) {\p_u}\ea $&$\ba{l} \nu\lo u{ {\p_u}}-t{\p_ t}\ro\\+{ {\p_v}}\
\text{ if }\lbd =0\ea $&$\ba{c}2\  \\\texttt{for}\\\text{ any}\ \lbd\ea$\\
\hline
    5.&$\ba{l}f^1=\lbd v^{\mu-1}e^{u},\\f^2=\s v^\mu
e^u\ea$&$\ba{l}
    D-{\p_u},\\ t {\partial_ t}+v {\p_v}
    -\mu  {\p_u}\ea$&$\ba{l}e^W
\ \texttt{if}\ \mu=1\  \&\\ H^a{\p_{ x_a}} - H^a_{x_a} v{\p_ v}\\
\text{if}\ m=2\ea$&$\ba{c}11\
\texttt{if}\\\mu=1\ea$\\
\hline 6.&$\ba{l}f^1=\lbd\ln v,\\f^2=\s
v^\frac{\mu+1}{2}\ea$&$\ba{l}\mu D-\frac{\mu+1}{2} t{\p_ t}
\\-\frac{1-\mu}{2}u {\p_u}-v {\p_v}\\
-\lbd t {\p_u},\ \Psi_0(x) {\p_u}\ea$&$\ba{l}x_a{\p_{ x_a}}+2u
{\p_u},\\u\p_u+2v\p_v+t\p_t\\+2\lbd t\p_u
\ \text{if}\ \s=0\ea$&$\ba{c}2\  \ea$\\
\hline
    7.&$\ba{l}f^1=\s v^{1-\mu},\\f^2=\lbd\ln v\ea$&$\ba{l}\mu D+\mu u\p_u-v\p_v\\-(1+\mu)t\p_t\\
- \frac{\lbd x^2}{2m}\p_u,\ \
\Psi(x)\p_u\ea$&$\ba{l}D+u\p_u-t\p_t\\ \texttt{if} \
\s=0\ea$&$\ba{c}2\  \ea$\\\hline
\end{tabular}

\newpage
\begin{center}{\bf
Table 10. Non-linearities with arbitrary parameters and non
extendible symmetries for equations (\ref{1.2}) with $a=0$}
\end{center}
   \begin{tabular}{|l|l|l|l|c|}
  \hline
No&Non-linearities&$\ba{c}\text{Condi-}\\\text{tions}\ea$
&Symmetries&$\ba{c}\text{AET}\\(\ref{eqv})\ea$\\
\hline
  1. & $\ba{l} f^1=\lambda u^{3\mu+1}v^\mu,\\f^2=\s u^{3\mu}v^{\mu+1}-\al u, \ea $&
   $\ba{l}\mu\neq 0,\\m=1,\\\al=-1\ea$ & $\ba{l}
   Q_7=4\mu t{\p_ t}\\-(\mu+1)u{\p_u}\\+(3\mu-1)v{\p_ v},\\ Q_2,Q_3\ea$& \\
   \cline{3-4}
   &$$&$\ba{l}\mu\neq 0,\\m=1,\\\al=1\ea$&$Q_4,Q_5,Q_7$&\\
    \hline
  2.& $\ba{l} f^1=\lambda u^{-2}v^{-1},\ea$&
    $\ba{l}m=1,\\\al=-1\ea$ & $\ba{l}e^{\va t}\lo{\p_ t}+\va v {\p_
v}\ro,\\ Q_2,\ Q_3\ea$& 11 if\\
    \cline{3-4}&$f^2=\s u^{-3}+\va v-\al u$&$\ba{l}m=1,\\\al=1\ea$&$\ba{l}e^{\va t}\lo{\p_ t}+\va v
    {\p_v}\ro,\\ Q_4,\ Q_5\ea$&$\lbd=0$ \\
   \hline
   3. & $\ba{l} f^1=\lbd v^{\mu+1},\\ f^2=\s v^{\mu-\nu+1}\ea$&$\mu\neq -1$&
  $\ba{l}\lo\mu-{2}\nu\ro D+\nu t
  {\p_ t}\\-v{\p_ v}-(\nu+1)u{\p_ u},\\
  \Psi_0(x){\p_u}\ea$ &2  \\
    \hline
 4.&$\ba{l}f^1=\lambda
v,\ f^2=e^{-v}\ea$&$\lbd\neq 0$&$\ba{l}
   2 D-
    t {\partial_ t}+u {\p_u}\\+ {\p_v}+\lbd t
{\p_u},\ \Psi_0(x) {\p_u}\ea$&2\\
    \hline
 5. &$\ba{l} f^1=\lambda e^{v},\ f^2=\sigma e^{v} \ea$&$\lbd\s\neq0$&$
\ba{l}D- {\p_v},\ \Psi_0 (x) {\p_u}\ea$&2
\\
\hline 6. &$
\begin{array}{l}
f^1=\lambda u^{\nu +1}e^{\mu \frac{v}{u}},\\
f^2=e^{\mu \frac{v}{u}}(\lambda v +\sigma
u)u^\nu\ea$&$\mu\lbd\neq0$&$\ba{l} \mu D-u{ {\p_v},}\\\nu D-u{
{\p_u}}-v{ {\p_v}}\ea$
&\\
 \hline
 7.&$ \ba{l} f^1=\mu \ln v,\\ f^2=\nu \ln v
    \ea$&$\nu\neq0$&$
\begin{array}{l}
\Psi_0 (x){\p_u}
,\\  D+u{ {\p_u}}+v{ {\p_v}}\\
{+}\left( \mu t-\frac {\nu}{2 m}x^2\right){\p_u}
\end{array}
$
&2\\
\hline
  8.&$f^1=0,\ f^2=\va \ln u$&$\va=\pm 1$&$\ba{l}D-t\p_t+u\p_u\\+\va
t\p_v,\  t\p_t+v\p_v, \\
\Phi(u,x)\p_v\ea$&$\ba{c}3,5,\\11\ea$\\\hline
    9. &$\ba{l} f^1=\va\lo\ln v-\kappa \ln u\ro u,
  \\f^2=\va\lo\ln v-\kappa \ln u\ro v \ea$ & $\ba{l}m\neq 2,\\
  \kappa\neq\frac{m+2}{m-2}\ea$ & $\ba{l}(1-\kappa)x_a{\p_{x_a}}\\+2\kappa v {\p_v}
+2u {\p_u},\\ e^{(1-\kappa)\va t} \lo u {\p_u}+ v {\p_v}\ro
  \ea$&$\ba{c}1\\\text{if}\\\kappa=1\ea$ \\
    \hline
  10. & $\ba{l} f^1=\va u\lo (m+2)\ln u\right.\\\left.+(2-m)\ln v\ro,
 \ea $& $\ba{l}m\neq 1,2\\
 \al=0\ea$ & $\ba{l}Q_1,\ x_a Q_1
 -x^2{\partial x_a},\\ e^{4\va t}\lo u
{\p_u}+v {\p_v}\ro\ea$& \\
\cline{3-4}
    &$\ba{l}f^2=\va v\lo (m+2)\ln u\right.\\\left.+(2-m)\ln v\ro-
\al u\ea$&$\ba{l}m=2,\\
\al=0\ea$&$\ba{l}H^a{\p_{x_a}} - H^a_{x_a}
v{\p_v},\\e^{4\va t}\lo u {\p_u}+v {\p_v}\ro\ea$&\\
\cline{3-4}
    &&$\ba{l}m=1,\\\al=1,\\\va=1\ea$&$\ba{l}Q_2,\ Q_3,\\ e^{4 t}\lo u {\p_u}+v {\p_v}\ro\ea$&\\
    \cline{3-4}
&&$\ba{l}m=1,\\\al=1,\\\va=-1\ea$&$\ba{l}Q_4,\ Q_5,\\ e^{-4 t}\lo u {\p_u}+v {\p_v}\ro\ea$&\\
     \hline
\end{tabular}

\newpage

\begin{center}{\bf
Table 10. Continued}
\end{center}
   \begin{tabular}{|l|l|l|l|c|}
  \hline
No&Non-linearities&Conditions&Symmetries&$\ba{c}\text{AET}\\(\ref{eqv})\ea$\\
\hline 11.&$\ba{l}f^1=\mu u\ln u,\\f^2=\mu v\ln u+\nu
v\ea$&$\mu\neq0$&$\ba{l} e^W\p_v, \\e^{\mu
t}(u\p_u+v\p_v)\ea$&11\\
    \hline
    12.&$\ba{l}f^1=\va v,\\
f^2=\lbd\frac{v^2}{u}+2\nu
v\ea$&$\ba{l}\lbd=\pm1,\\\s=\mp\nu^2\ea$&$\ba{l} Q_8=e^{\nu
t}(\lbd(u\p_u+v\p_v)\\+\nu u\p_v),\ e^{\nu
t}u\p_v+tQ_8\ea$&\\
\cline{3-5} &$\ba{l}+\s u\ln
u\\\\\ea$&$\ba{l}\lbd\neq0,\\\nu^2+\lbd\s\\=1\ea$&$\ba{l}
X_\pm=e^{\nu\pm1}(\lbd(u\p_u\\+v\p_v)\\+(\nu\pm1)u\p_v)\ea$&$\ba{c}1\ \text{if}\\\s=0\ea$\\
    \cline{3-5}
&&$\ba{l}\lbd\neq0,\\\nu^2+\lbd\s\\=-1\ea$&$\ba{l}e^{\nu
t}(\lbd\cos
t(u\p_u+v\p_v)\\+(\nu\cos t-\sin t)u\p_v),\\
e^{\nu t}(\lbd \sin t(u\p_u+v\p_v)\\+(\nu\sin t+\cos t)u\p_v)\ea$&\\
 \hline

       \end{tabular}
\vspace{3mm}

The classification of systems of coupled reaction-diffusion
equations has been ended.

\section{Discussion}

We complete the group classification of systems of
reaction-diffusion equations started in papers \cite{N1},
\cite{N2}, where systems with a diagonal and square diffusion
matrix are studied.

The case of triangular diffusion matrix considered in the present
paper appears to be rather complicated mainly due to very large
number of versions with different symmetries. Indeed, we indicate
54 non-equivalent equations (\ref{1.2}) with an invertible
triangular diffusion matrix which are presented in Tables 2-6 and
68 equations with a nilpotent diffusion matrix which are present
in Tables 8-10 and partly in Tables 2, 3.  In addition,  14
classes of  equations including first derivatives w.r.t. variables
$x_{\nu} $ are collected in Table 7.

In Tables 2, 3 equations defined up to arbitrary functions are
presented. In the third column of Table 1 we present the main
symmetries of the related equations (\ref{1.2}) which do not admit
extended symmetries.

In Table 3 we present equations (\ref{1.2}) which admit additional
equivalence transformations (AET) (\ref{eqv}) (indicated in the
fourth column) and possible extended symmetries indicated if
necessary in the third column after the symbol $\&$.

Tables 4-6 includes the results of classification of equations
(\ref{1.2}) which are defined up to arbitrary parameters.

The items of Tables 2 and 3 marked by the asterisks are related
for equations (\ref{1.2}) with $a\neq0$ and $a=0$ as well.

Tables 7-10 present the results of group classification of
equations (\ref{1.2}) with the nilpotent diffusion matrix only.

Note that we did not consider linear equations whose group classification 
is a rather trivial problem.

Among the classified equations there are only seven of them being
invariant w.r.t. the Galilei transformations. In accordance with
Item 2 of Table 3 the general form of Galilei-invariant equations
(\ref{1.2}) is
 \be\label{an1}\ba{l}u_t-a\Delta u=uF_1,\\v_t-\Delta u-a\Delta v=uF_2-vF_1\ea\ee
where $F_1$ and $F_2$ are arbitrary functions of variable
$\xi=ue^{\frac{v}{u}}$. In Items 3, 5 of Table 4 and Items 1-4 of
Table 6 all functions $F_1$, $F_2$ are presented which correspond
to various possible extensions of the Galilean symmetry. The only
system of equations (\ref{1.2}) which admits the extended Galilei
group including the dilatation and conformal transformations is
given by formula (\ref{n06}).

Thus we end the group classification of systems of coupled
reaction-diffusion equations (\ref{1.1}) started in paper
\cite{danil} and continued with varying success in
\cite{nikwil2}-\cite{chern3} and \cite{N1}, \cite{N2}. The number
of non-equivalent equations of this type appears to be enormously
large ($>300$). Nevertheless, using the approach presented in
\cite{N1} it was possible to make an effective classification of
pairs of coupled reaction-diffusion equations with general
diffusion matrix. Moreover, we classify the equations with
arbitrary number of independent variables.

We notice that group analysis of reaction-diffusion equations is
being intensively developed in many lines including equations with
arbitrary elements depending on $t,x, u, u_t, u_x,\cdots$, refer
to \cite{pop2} for a survey. Thus our analysis of systems of
diffusion equations is nothing but a part of general study of
diffusion models which is currently rather popular.


\begin{thebibliography}{99}
\bibitem{N1} A. G. Nikitin. Group classification of systems of non-linear reaction-diffusion equations
with general diffusion matrix. I. Generalised Ginzburg-Landau
equations, math-ph/0411027, 2004.
\bibitem{N2}A. G. Nikitin,  Group classification of systems of non-linear
reaction-diffusion equations with general diffusion matrix. II.
Diadonal diffusion matrix, math-ph/0411028, 2004.
\bibitem{danil} Yu. A. Danilov, {\it Group analysis of the Turing systems
and of its analogues}, Preprint of Kurchatov Institute for Atomic
Energy IAE-3287/1, 1980.
\bibitem{nikwil2}A. G. Nikitin and R. Wiltshire,
in: {\it Symmetries in Nonlinear Mathematical Physics, Proc. of
the Third Int. Conf. , Kiev, July 12-18, 1999, Ed. A.M.
Samoilenko} ( Inst. of Mathematics of Nat. Acad. Sci. of Ukraine,
Kiev, pp. 47-59, 2000).
\bibitem{chern1} R. M. Cherniha and J. R. King, J. Phys. A {\bf
33}, 267-282, 2000.
\bibitem{chern2} R. M. Cherniha and J. R. King, J. Phys. A {\bf
33}, 7839-7841, 2000.
\bibitem{nikwil1}A. G. Nikitin and R. Wiltshire, J. Math. Phys. {\bf
42} 1667, 2001.
\bibitem{pop}A. G. Nikitin and R. O. Popovych, Ukr. Mat. Zhurn. \textbf {53},
    1053-1060, 2001.
\bibitem{chern3} R. M. Cherniha and J. R. King, J. Phys. A {\bf
36}, 405-425, 2002.
\bibitem{pop2}R.O. Popovych and N.M. Ivanova, J. Phys. A {\bf 37}, , V.37, 7547-7565,
2004 (see also math-ph/0306035).

\end{thebibliography}
\end{document}